\documentclass[twoside]{ilcws07}
\usepackage[latin1]{inputenc}
\usepackage[dvips]{graphicx,epsfig,color}
\usepackage{wrapfig,rotating}
\usepackage{amssymb,amsmath,array}
\newcommand{\whizard}{\texttt{WHIZARD}}
\newcommand{\GeV}{{\ensuremath\rm GeV}}
\newcommand{\TeV}{{\ensuremath\rm TeV}}
\newcommand{\chp}{\tilde{\chi}^+}
\newcommand{\chm}{\tilde{\chi}^-}
\newcommand{\om}{\texttt{O'Mega}}
\newcommand{\ME}{\mathcal{M}}
\newcommand{\ab}{{\ensuremath\rm ab}}

\begin{document}
\bibliographystyle{unsrt}
\title{
Monte Carlo Simulations \\for NLO Chargino Production at the ILC }
\author{Tania Robens
\thanks{Work supported by DFG SFB/TR9
"Computational Particle Physics" and German Helmholtz Association, Grant VH-NG-005.}
\vspace{.3cm}\\
RWTH Aachen - Institut f\"ur Theoretische Physik E \\
52056 Aachen - Germany
}

\maketitle

\begin{abstract}
 We present an extension of the Monte Carlo Event Generator \whizard~
 which includes chargino production at the ILC at NLO. We include photons using both a fixed order and a resummation approach. In the latter, leading higher order corrections are automatically included. We present results for cross
 sections and event generation for both methods \cite{slides}.This is an updated version of the results presented in \cite{Kilian:2006bg}. 
\end{abstract}
\centerline{\texttt{PITHA 07/10, SFB/CPP-07-51}}
\section{Introduction}
In many GUT models, the masses of charginos tend to be near the lower
edge of the superpartner spectrum, and  they can be pair-produced at a first-phase ILC with c.m.\ energy of
$500\;\GeV$. The precise measurement of their parameters (masses,
mixings, and couplings) is a key for uncovering 
the fundamental properties of the
MSSM~\cite{Aguilar-Saavedra:2005pw}. Regarding the
experimental precision at the ILC, off-shell kinematics for the signal process,
the reducible and irreducible backgrounds~\cite{Hagiwara:2005wg}, and NLO corrections need to be included.  We here present the inclusion of NLO chargino production where corrections can be in the percent regime.
\section{Chargino production at LO and NLO}


The total fixed-order NLO cross section is given by
\begin{equation}\sigma_{\rm tot}(s,m_e^2) = \sigma_{\rm Born}(s) + 
  \sigma_{\rm v+s}(s,\Delta E_\gamma,m_e^2) + 
  \sigma_{\rm 2\rightarrow 3}(s,\Delta E_\gamma,m_e^2),
\end{equation}
where $s$ is the cm energy, $m_{e}$ the electron mass, and
$\Delta\,E_{\gamma}$ the soft photon energy cut dividing the photon
phase space.  
The 'virtual' 
contribution $\sigma_\text{v}$ is the interference of the one-loop corrections
\cite{Fritzsche:2004nf} with the Born term. The collinear and infrared
singularities are regulated by $m_e$ and the photon mass $\lambda$, respectively. 
The dependence on $\lambda$ is eliminated by
adding the soft real photon contribution $\sigma_{\rm s}
\,=\,f_{\rm soft}\,\sigma_{\rm Born}(s)$ with a universal soft factor 
$f_{\rm soft}(\frac{\Delta E_\gamma}{\lambda})$
\cite{Denner:1991kt}. We break the `hard' contribution 
$\sigma_{\rm 2\rightarrow 3}(s,\Delta E_\gamma,m_e^2)$, i.e., the
real-radiation process $e^-e^+\rightarrow\chm_i\chp_j\gamma$,
into a
collinear and a non-collinear part, separated at a photon
acollinearity angle $\Delta\theta_\gamma$ relative to the incoming
electron or positron. The collinear part is approximated by convoluting the Born cross section with a
structure function $f(x;\Delta\theta_\gamma,\frac{m_e^2}{s})$
\cite{Bohm:1993qx}. The non-collinear part is generated
explicitely.
 
The total fixed order cross section 
is implemented in the multi-purpose event generator
\om/\whizard~\cite{Moretti:2001zz,Kilian:2007gr} using 
a `user-defined' structure 
function and an effective matrix
element $|\ME_\text{eff}|^2$ which contains the
Born part, the soft-photon factor and the Born-1 loop interference
term.  
\begin{figure}
\centering
\includegraphics[width=.45\textwidth]{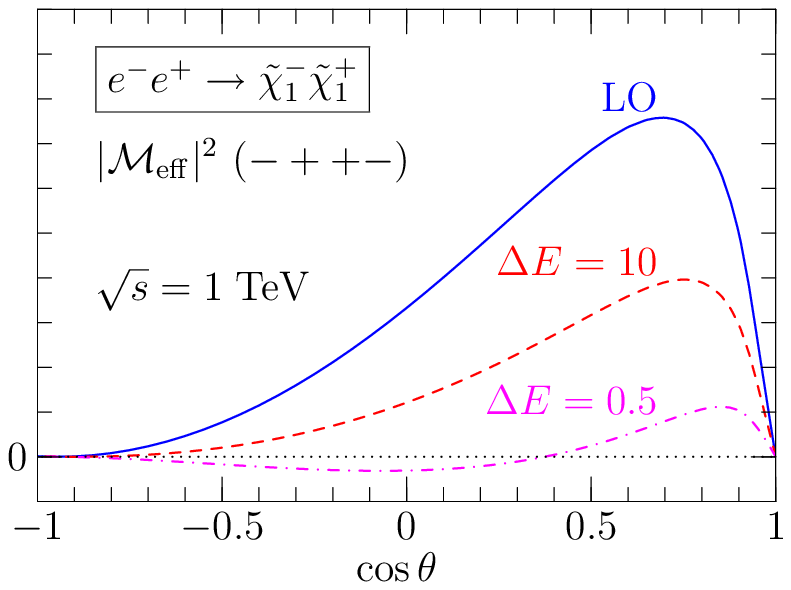} \quad
\includegraphics[width=.45\textwidth]{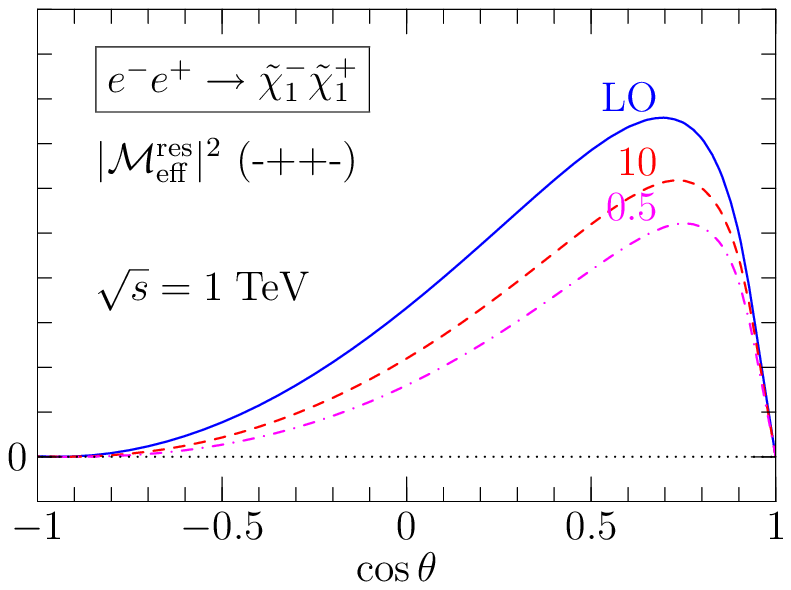}
\caption{{\small $\theta$ -dependence of effective squared matrix element 
  ($\sqrt{s}=1\;\TeV$).Left figure: fixed order effective matrix element; right
  figure: effective matrix element with
  the one-photon ISR part subtracted.  Solid line: Born term; dashed: including virtual
  and soft contributions for $\Delta E_\gamma=10\;\GeV$; dotted: same with
  $\Delta E_\gamma=0.5\;\GeV$.  
  $\Delta\theta_\gamma=1^\circ$.}}
\label{fig:Meff}
\end{figure}
In the soft-photon region this approach runs into
the problem of negative event
weights~\cite{Kleiss:1989de}: for some values of $\theta$,  
the $2\to 2$ part of the NLO-corrected
squared matrix element is positive definite by itself only if $\Delta
E_\gamma$ is sufficiently large, cf Fig.~\ref{fig:Meff}.
To still obtain unweighted event samples, an ad-hoc approach is to simply drop events with negative events before
proceeding further.

Negative event weights can be avoided by
resumming
higher-order initial
radiation using an exponentiated structure function 
$f_\text{ISR}$~\cite{Skrzypek:1990qs}. In order to avoid double-counting in the 
combination of the ISR-resummed LO result with the additional NLO 
contributions \cite{Fritzsche:2004nf}, we have subtract from the effective
squared matrix element the soft and virtual photonic contributions that have already been
accounted for in $\sigma_\text{s+v}$.
This defines $|\ME^{\text{res}}_\text{eff}|^2 \, = \,
|\ME_\text{eff}|^{2}- 2f_\text{soft,ISR}  \,|\ME_\text{Born}|^2$ which is positive for even low $\Delta E_{\gamma}$ cuts for all values of $\theta$ (cf Fig. \ref{fig:Meff}), such that unweighting of generated events and
realistic simulation at NLO are now possible in all regions of
phase-space.
 Convoluting
this with the resummed ISR structure function for each incoming beam,
we obtain a modified $2\to 2$ part of the total cross section which
also includes soft and collinear photonic 
corrections to the Born/one-loop interference. This differs from the standard treatment in the literature (cf eg. \cite{Fritzsche:2004nf}) where higher order photon contributions are combined with the Born term only (``Born+''). The complete result
contains the hard non-collinear $2\to 3$ part convoluted with
the ISR structure function:
\begin{equation}
\sigma_{\text{res,+}}=\int^{\Delta (E,\theta)} \,dx_{i}\,d\Gamma_{2}\,f_{\text{ISR}}^{(e^{+})}(x_{1})f_{\text{ISR}}^{(e^{-})}(x_{2})  |\ME^{\text{res}}_{\text{eff}}|^{2}+\int_{\Delta (E,\theta)} dx_{i}\,d\Gamma_{3}\,f_{\text{ISR}}^{(e^{+})}(x_{1})f_{\text{ISR}}^{(e^{-})}(x_{2})|\ME^{2\rightarrow\,3}|^{2}
\end{equation}


\section{Results}
\begin{figure}
\centering
  \includegraphics[height=0.39\textwidth,width =0.8\textwidth]{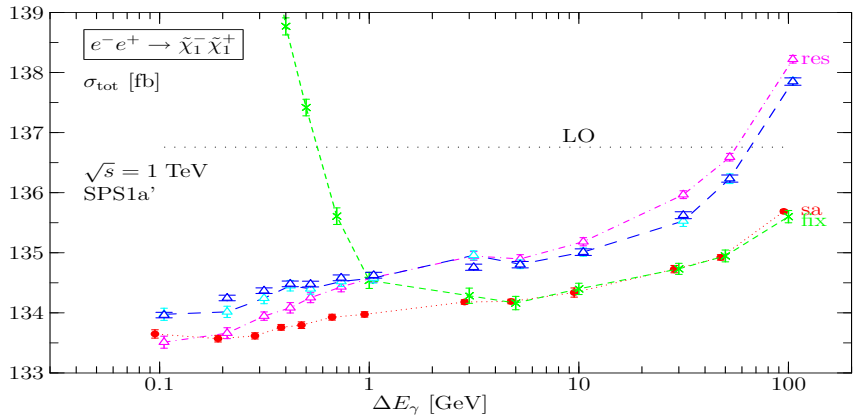}
  \caption{\label{fig:edep} {\small Total cross section dependence on $\Delta
    E_\gamma$: {\rm `sa'}
    (dotted) = fixed-order semianalytic result; {\rm `fix'} (dashed) = fixed-order
    Monte-Carlo result; {\rm `res'}
    (long-dashed) = ISR-resummed Monte-Carlo result; (dash-dotted) = same but resummation applied only to
    the $2\to 2$ part. $\Delta\theta_\gamma=1^\circ$.
    LO: Born cross section.}}
\label{fig:edep}
\end{figure}
 Fig.~\ref{fig:edep} compares the $\Delta E_{\gamma}$ dependence of
 the numerical results from 
the semianalytic fixed-order calculation with the Monte-Carlo
integration in the fixed-order and in the resummation schemes. The fixed-order Monte-Carlo result agrees with the semianalytic
result as long as the cutoff is greater
than a few $\GeV$ but departs from it for smaller cutoff values
because here, in some parts of phase 
space, $|\ME_{\rm eff}|^{2}\,<\,0$ is set to zero. The semianalytic
fixed-order result is not exactly 
cutoff-independent, but exhibits 
a slight rise of the calculated cross section with increasing cutoff
(breakdown of the soft approximation). For $\Delta E_\gamma=1\;\GeV$
($10\;\GeV$) the shift is about 
$2\,\text{permil}$ ($5\,\text{permil}$) of the total cross section. The fully resummed result  shows an increase of about
$5\,\text{permil}$ of the total cross section with respect to the
fixed-order result which stays roughly constant until $\Delta
E_\gamma>10\;\GeV$.  This is due to higher-order photon radiation.

\begin{figure}[t]
\centering
  \includegraphics[height=.35\textwidth,width=.95\textwidth]{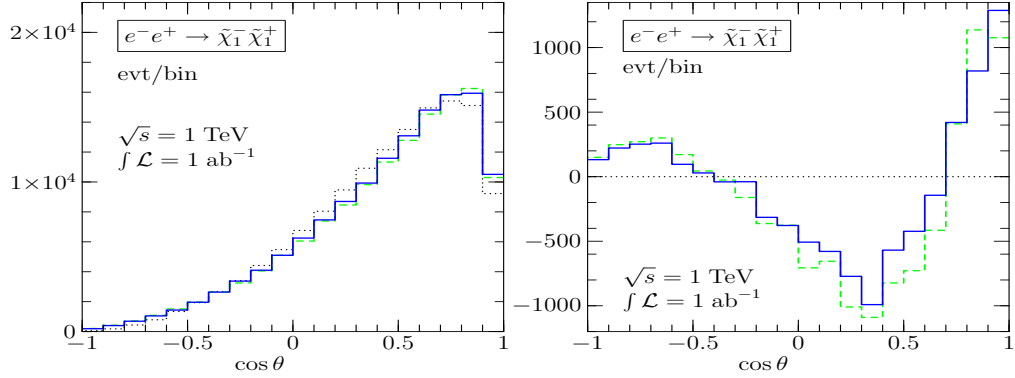}
\vspace{-3mm}
  \caption{{\small Polar scattering angle distribution for an integrated
    luminosity of $1\;\ab^{-1}$ at $\sqrt{s}=1\;\TeV$. Left: total
    number of events per bin; right: difference w.r.t.\ the Born
    distribution.  LO (dotted) = Born cross section without
    ISR; fix (dashed) = fixed-order approach; res (full)
    = resummation approach.  Cutoffs: $\Delta E_\gamma=3\;\GeV$ and
    $\Delta\theta_\gamma=1^\circ$.}}
\label{fig:histth}
\end{figure}

 In
Fig.~\ref{fig:histth} we show the binned distribution of the chargino
production angle obtained using a sample of unweighted events. 
It demonstrates that NLO corrections (which, for total cross sections, are in the percent regime and can reach $20\%$ at the threshold) are important and cannot be
accounted for by  a constant K factor. 
\vspace{2mm}
\begin{figure}
\centering
  \includegraphics[height=0.3\textwidth,width =0.6\textwidth]{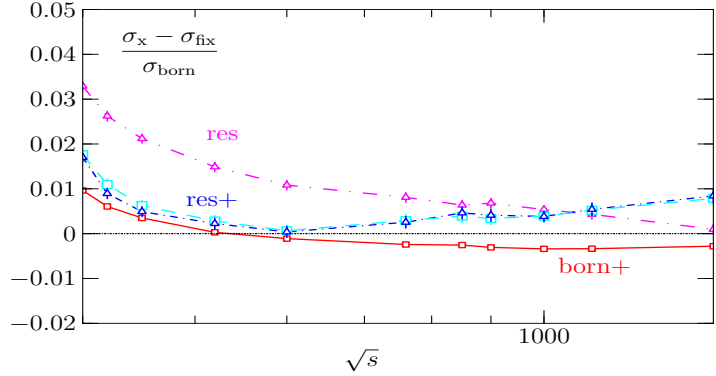}
\vspace{2mm}
  \caption{{\small  Relative higher-order effects for different methods: (magenta, long dash dotted) = $\sigma_{\text{res}}$, (blue/ cyan and dash-dotted/ dashed) =  $\sigma_{\text{res}+}$, and (red, solid) = $\sigma_{\text{Born}+}$  vs $\sigma_\text{Born}$}.    }
\label{fig:secoeff}
\end{figure}
Figure \ref{fig:secoeff} shows the magnitude of second and higher order photonic effects in different schemes. Resummation effects are clearly in the percent regime and cannot be neglected. For $\sqrt{s}\,>\,500\GeV$, the convolution of the interference term with $f_\text{ISR}$ additionally changes the sign of the higher order corrections.
 For more details, cf.~\cite{Kilian:2006cj,Robens:2006np}.
\section{Conclusions}
We have implemented NLO corrections
into the event generator \whizard~for chargino pair-production at the
ILC with several approaches for the inclusion of photon radiation. 
A careful analysis of the dependence on the cuts
$\Delta\,E_{\gamma},\,\Delta\,\theta$  reveals 
uncertainties related to higher-order radiation and breakdown of the
soft or collinear approximations. To carefully choose the
resummation method and cutoffs will be critical for a truly precise
analysis of real ILC data.The version of the program resumming
photons allows to get rid of negative event weights,
accounts for all yet known higher-order effects, allows for 
cutoffs small enough that soft- and collinear-approximation artefacts
are negligible, and explicitly generates photons where they can be
resolved experimentally. Corrections for the decays of charginos and
non-factorizing corrections are in the line of future work. 

\begin{footnotesize}


\bibliographystyle{unsrt}
\bibliography{ilcws07}
\end{footnotesize}
\end{document}